\documentclass[12pt]{article}
\usepackage{amsfonts}
\usepackage{eurosym}
\usepackage{amssymb}
\usepackage{amsmath}
\usepackage{cite}

\newtheorem{theorem}{Theorem}

\begin{document}

\title{Projective collineations of decomposable spacetimes generated by the
Lie point symmetries of geodesic equations}
\author{Andronikos Paliathanasis\thanks{%
Email: anpaliat@phys.uoa.gr} \\
{\ \textit{Institute of Systems Science, Durban University of Technology }}\\
{\ \textit{PO Box 1334, Durban 4000, Republic of South Africa}} \\
{\textit{Instituto de Ciencias F\'{\i}sicas y Matem\'{a}ticas,}}\\
{\ \textit{Universidad Austral de Chile, Valdivia, Chile}}\\
}
\maketitle

\begin{abstract}
We investigate the relation of the Lie point symmetries for the geodesic
equations with the collineations of decomposable spacetimes. We review
previous results in the literature on the Lie point symmetries of the
geodesic equations and we follow a previous proposed geometric construction
approach for the symmetries of differential equations. In this study we
prove that the projective collineations of a $\left( n+1\right) \,$%
-dimensional decomposable Riemannian space are the Lie point symmetries for
geodesic equations of the $n$-dimensional subspace. We demonstrate the
application of our results with the presentation of applications. \newline
\newline

\end{abstract}

\section{Introduction}

\label{sec1} The theory of Lie point symmetries for the study of
differential equations provides a systematic approach for the investigation
of similarity solutions for nonlinear dynamical systems. This powerful
method established by Sophus Lie at the end of the 19th century \cite%
{lie1,lie2,lie3}. The novelty of S. Lie approach is that he moved the
resolution of differential equations from a study of continuous
transformations within the context of geometry. The geometric background
brought with it the constriction of mapping from a point to a point through
transformations. The symmetry analysis has been widely studied in the
literature. The simplicity on the steps of the theory and the unexpectedly
number of new results which were found the last decades on nonlinear
systems, in all areas of applied mathematics \cite%
{od1,od3,el3,el4,hp1,sw3,an2,an3,an4,mm1,mm2,mm3,mm4,mm5}, established the
Lie symmetry analysis as one of the most important methods for the study of
nonlinear differential equations. Indeed, there are many important results
in real world problems which followed by Lie symmetry analysis. An
interesting discussion on the Lie symmetries in epidemiology is presented in
\cite{nn1}, while the analysis of the differential equations of financial
mathematics with the use of Lie symmetries provided a new point of view for
this area \cite{nn2,nn3}. Moreover, an application of Lie's theory in the
dynamics of meteorology is discussed in \cite{nn4}.

One of the most important direct result of the Lie point symmetry analysis
is the linearization of second-order ordinary differential equations. Sophus
Lie on his work proved that for every second-order ordinary differential
equation which admits eight Lie point symmetries there exists a
transformation in which the given differential equation can be written in
the equivalent form of the free particle \cite{lie1}. Other linearization
criteria established in the literature for higher-order differential
equations and for partial differential equations \cite{ln1,ln2,ln3,ln4}. In
addition, it was found that differential which admit the same symmetries has
similar properties, Ovsiannikov carried \cite{Ovsi} the first group
classification problem and demonstrate the construction of the
one-dimensional optimal system for the Lie algebra.

The concept of symmetry is essential in essential in geometry and on
physical theories such is general relativity and cosmology. The
determination of symmetries in Riemannian manifolds is essential for the
derivation of new exact solutions of Einstein's field equations \cite%
{m1,m2,m3,m4,m5,m6}. Moreover, the admitted conformal algebra of a manifold
is used to perform a classification of Riemannian spaces \cite{Frances,Johar}%
. A special class of conformal symmetries are the isometries, which has been
used in numerous applications, such as to simplify the gravitational field
equations not only in General Relativity \cite{Geroch1,Geroch2} but in other
theories such is the Einstein-Skyrme model \cite{es1,es2,es3} or to study
the geometric properties of the background space, such is the decomposition
property of the metric tensor \cite{ac1}.

In Riemannian spaces, the Levi-Civita connection inherent a system of
differential equations defined by the paths of the connection. In terms of
differential geometry, these curves remain invariant under the projective
collineations. Consequently, there should be a connection between the Lie
point symmetries of differential equations with the space collineations.
This was the subject of study from different scientific groups, such is that
of Katzin et al. \cite{k1,k2,k3}, Aminova \cite{ami2} and others \cite%
{b1,b2,b3,b4,b5}. In \cite{ts1}, Tsamparlis and Paliathanasis found a
systematic method for the construction of the Lie point symmetries for the
geodesic equations of Riemannian spacetimes by using the elements of the
projective algebra for the background space. Furthermore, a similar
geometric construction approach was applied for the construction of Noether
symmetries \cite{ts2}. For the geodesic Lagrangian Noether symmetries were
found to be constructed by the elements of the homothetic algebra for the
metric tensor. Some recent extension of this approach on holonomic and
unholonomic systems can be found in \cite{tsa1,tsa2,tsa3}.

In this piece of work, previous results on the relation between Lie point
and Noether point symmetries for the geodesic equations with the
collineations of the background space are reviewed. In Section \ref{sec2} we
present the basic properties and definitions for point transformations and
invariant functions. That definitions are extended in the case of geometric
objects in Section \ref{sec3}. Moreover, the concept of geometric
collineations is discussed, where we focus on the collineations of the
metric tensor and of the connection. The Lie point symmetries for
differential equations are discussed in Section \ref{sec4}. In addition,
Noether's two theorems are presented. In Section \ref{sec5} we present two
Theorems which gives the explicitly relation of the point symmetries for the
geodesic equations with the collineations. Furthermore, in Section \ref{sec6}
we present the new results of this study where we show that the Lie point
symmetries for the geodesic equations of $n$-dimensional Riemannian manifold
form the projective algebra for a $\left( n+1\right) $-dimensional
decomposable Riemannian space. Indeed, we prove by using the symmetries of
differential equations that the projective algebra of the $\left( n+1\right)
$-dimensional decomposable space can be constructed by the projective
collineations of the $n$-dimensional subspace. Until now collineations have
been used mainly for the construction of the symmetries of differential
equations. However, in this study we focus on the inverse approach. Finally
in Section \ref{sec7}, we summarize our results and we draw our conclusions.

\section{Point transformations and invariant functions}

\label{sec2}

Consider $M~$to be a manifold of class$~C^{p}~$with$~p\succeq 2$ and $U$ to
be a neighborhood in $M$. For the two points $P,Q\in U~$with coordinates $%
\left( x_{P},y_{P}\right) $ and \ $\left( x_{Q}^{\prime },y_{Q}^{\prime
}\right) $ respectively, the point transformation on $U$ which drags point $%
P $ on $Q$\ is defined as%
\begin{equation}
x_{Q}^{\prime }=x^{\prime }\left( x_{P},y_{P}\right) ~~,~~y_{Q}^{\prime
}=y^{\prime }\left( x_{P},y_{P}\right)
\end{equation}%
in which functions $x^{\prime }\left( x,y\right) ,~y^{\prime }\left(
x,y\right) ~$are $C^{p-1}$ and $\det \left\vert \frac{\partial \left(
x^{\prime },y^{\prime }\right) }{\partial \left( x,y\right) }\right\vert
\neq 0.$The later condition is necessary in order the functions $x^{\prime
}\left( x,y\right) ,~y^{\prime }\left( x,y\right) $ to be independent,
otherwise the point transformation degenerates. When the point
transformation depends on one parameter$~\varepsilon ,$ is called one
parameter point transformation \cite{StephaniB}.

Therefore, if $\Phi $ is the one parameter point transformation $\Phi
:P\rightarrow Q$, then%
\begin{equation}
x_{Q}^{\prime }=x^{\prime }\left( x_{P},y_{P};\varepsilon \right)
~~,~~y_{Q}^{\prime }=y^{\prime }\left( x_{P},y_{P};\varepsilon \right) .
\label{PT.0A}
\end{equation}%
By definition, $\Phi $ it admits the group properties, that is, which means
that for every function $\Phi $ ant on $P\in U,~\Phi :P\rightarrow P^{\prime
}$ the resulting point $P^{\prime }\in U$.

Moreover, the identity transformation exists, $\Phi _{I}:P\rightarrow P$ $,$
there exists the associativity, and the inverse element. A group of one
parameter point transformation defines a family of curves in $M,$ which are
parametrized by the parameter $\varepsilon $ and are called the orbits of
the group of transformations.

When parameter $\varepsilon $ is infinitesimal we can define the tangent
vector a the point $P$ as
\begin{equation}
X_{P}=\frac{\partial x^{\prime }}{\partial \varepsilon }|_{\varepsilon
\rightarrow 0}\partial _{x}|_{P}+\frac{\partial y^{\prime }}{\partial
\varepsilon }|_{\varepsilon \rightarrow 0}\partial _{y}|_{P}.
\end{equation}%
where now for $\varepsilon ^{2}\rightarrow 0$ it follows%
\begin{equation}
x^{\prime }=x+\varepsilon \xi _{P}~~,~~y^{\prime }=y+\varepsilon \eta _{P}
\label{PT.02}
\end{equation}%
where $\xi _{P}=\frac{\partial x^{\prime }}{\partial \varepsilon }%
|_{\varepsilon \rightarrow 0}~~,$~~$\eta =\frac{\partial y^{\prime }}{%
\partial \varepsilon }|_{\varepsilon \rightarrow 0}.$

The vector field $X_{P}$ is called the generator of the infinitesimal point
transformation (\ref{PT.02}). The infinitesimal transformation is the local
transformation for the one parameter point transformation (\ref{PT.0A}). The
novelty of the point transformations is that for a given local infinitesimal
point transformation there can be always defined the global point
transformations and vice verse. Indeed, the corresponding point
transformation of an infinitesimal transformation can be found by the
derivation of integral curve for the infinitesimal generator $X_{P}$.

Consider the function $F\left( x,y\right) $ in manifold $M.$ Hence, under
the action of point transformation $\Phi $ it follows $\Phi :F\left(
x,y\right) \rightarrow F^{\prime }\left( x^{\prime }\left( x,y;\varepsilon
\right) ,y^{\prime }\left( x,y;\varepsilon \right) \right) $. Hence,
function $F$ is invariant under the point transformation $\Phi $ if and only
if $F\left( x,y\right) =F^{\prime }\left( x^{\prime }\left( x,y;\varepsilon
\right) ,y^{\prime }\left( x,y;\varepsilon \right) \right) $ or $F\left(
x,y\right) =\lambda \left( F^{\prime }\left( x^{\prime }\left(
x,y;\varepsilon \right) ,y^{\prime }\left( x,y;\varepsilon \right) \right)
\right) $ with $F\left( x,y\right) =0,$ where $\lambda $ is a function, at
all points where the one parameter point transformation acts

Equivalently, function, or equation, $F\left( x,y\right) =0,$ is invariant
under the action of an infinitesimal point transformation if the following
condition is true%
\begin{equation}
X\left( F\right) =0,  \label{InF.01}
\end{equation}%
that is
\begin{equation}
\xi \frac{\partial F}{\partial x}+\eta \frac{\partial F}{\partial y}=0.
\label{InF.02}
\end{equation}

The invariant functions of the vector field $X$ are derived from the
Lagrange $\frac{dx}{\xi \left( x,y\right) }=\frac{dy}{\eta \left( x,y\right)
}$. $\ $Hence every function of the form $F\left( x,y\right) =F\left(
W\right) $, in which $W$ is the zero-order invariant $dW=\frac{dx}{\xi
\left( x,y\right) }-\frac{dy}{\eta \left( x,y\right) }$, is invariant under
the infinitesimal transformation with generator the vector field $X$. \

Furthermore, the vector field $X$ is called Lie point symmetry for the
function $F\left( x,y\right) $. Since, the one parameter point
transformations form a group, then the infinitesimal generators $\left\{
X_{1},X_{2},...,X_{N}\right\} $ which are symmetry vectors for function $%
F\left( x,y\right) $ form a Lie algebra.

Until this point we have discussed the Lie point symmetries for functions.
In the following, we focus with the application of point transformations on
differential equations and on geometric objects in Riemannian manifolds.

\section{Geometric objects and collineations}

\label{sec3}

Let $\Omega $ be a geometric object of class $r$ on the $C^{p}$ manifold $M$%
, with $r\leq p$ . Then $\Omega $ is well defined on a every point $P$, that
is $\Omega ^{a}=\Omega ^{a}\left( x,y\right) $. Additionally, under a
coordinate transformation $\left( x,y\right) =J^{i}\left( x,y\right) $ the
new components $\Omega ^{a^{\prime }}$ of the object in the new coordinates $%
\left\{ x^{\prime }y^{\prime }\right\} $ are represented as well determined
functions of class $r^{\prime }=p-r$ of the old components $\Omega ^{a}$ in
the old coordinates $\left\{ x,y\right\} $, of the functions $J^{i}$ and of
their s-th derivatives $\left( 1\leq s\leq p\right) $, that is, the new
components $\Omega ^{i^{\prime }}$ of the object can be represented by
equations of the form $\Omega ^{a^{\prime }}=\Phi ^{a}\left( \Omega
^{k},x,y,x^{\prime },y^{\prime }\right) ,$ where functions $\Phi ^{a}$ have
the group properties. The transformation law, i.e. functions $\Phi ^{a}$
characterize the geometric object. In the following we are interesting on
differential geometric objects in which
\begin{equation}
\Phi ^{a}\left( \Omega ,x^{k},x^{k^{\prime }}\right) =J_{b}^{a}\left(
x,y,x^{\prime },y^{\prime }\right) \Omega ^{b}+C\left( x,y,x^{\prime
},y^{\prime }\right) .  \label{ls.01}
\end{equation}%
and on the special case with $C\left( x^{k},x^{k^{\prime }}\right) $, where $%
\Omega ^{a}$ are known as tensors and $J_{b}^{a}\left( x,y,x^{\prime
},y^{\prime }\right) $ is the Jacobian tensor.

Consider now the infinitesimal transformation (\ref{PT.02}) with generator
the vector field $X$. Then, under the map $\Phi $ the geometric object
differs as%
\begin{equation}
\mathcal{L}_{X}\Omega =\lim_{\varepsilon \rightarrow 0}\frac{1}{\varepsilon }%
\left[ \Phi \left( \Omega \right) -\Omega \right] .  \label{ls.02}
\end{equation}%
Operator $\mathcal{L}_{X}\Omega $ is the Lie derivative with respect to the
vector field $X$ on the geometric object $\Omega $. In terms of coordinates
the definition of Lie derivative depends on the transformation law $\Phi
^{a} $.

Indeed, for a function $F\left( x,y\right) $, the Lie derivative is defined
as $\mathcal{L}_{X}F=X\left( F\right) $. Furthermore, for tensor field~$T$
of rank $\left( r,s\right) $ is defined as \cite{ky}
\begin{eqnarray}
\mathcal{L}_{X}T_{~~j_{i}...j_{s}}^{i_{1}...i_{r}}
&=&X^{k}T_{~~j_{i}...j_{s,k}}^{i_{1}...i_{r}}-T_{~~j_{i}...j_{s}}^{m...i_{r}}X_{,m}^{i_{1}}-T_{~~j_{i}...j_{s}}^{i_{1}m...i_{r}}X_{m}^{i_{2}}+...
\notag \\
&&...+T_{~~m...j_{s}}^{i_{1}...i_{r}}X_{,j_{1}}^{m}+T_{~~j_{i}m...j_{s}}^{i_{1}...i_{r}}X_{j_{2}}^{m}+....
\label{ls.03}
\end{eqnarray}%
where the Einstein summation convention is considered.

On the other hand, the connection coefficients have different transformation
law from tensor fields, that means that the Lie derivative is defined
different. Specifically, the definition of~$\mathcal{L}_{X}\Gamma _{jk}^{i}$
is
\begin{equation}
\mathcal{L}_{X}\Gamma _{jk}^{i}=X_{,jk}^{i}+\Gamma
_{jk,r}^{i}X^{r}-X_{,r}^{i}\Gamma _{jk}^{r}+X_{,j}^{s}\Gamma
_{sk}^{i}+X_{,k}^{s}\Gamma _{js}^{i}.  \label{ls.04}
\end{equation}%
When we have a symmetric connection $\Gamma _{jk}^{i}=\Gamma _{kj}^{i},$the
latter expression can be written in the equivalent form%
\begin{equation}
\mathcal{L}_{X}\Gamma _{jk}^{i}=X_{;jk}^{i}-R_{jkl}^{i}\xi ^{l},
\label{ls05}
\end{equation}%
in which $R_{jkl}^{i}$ is the curvature tensor and the semicolon~\thinspace $%
";"$ means covariant derivative.

Previously, the invariant functions under point transformations were
defined. In the folloiwng lines, we can easily extend the same definition in
the case of geometric objects. Therefore, it will be said that a geometric
object $\Omega $ is invariant under the action of the one parameter point
transformation (\ref{PT.0A}) , $\Phi :\Omega \rightarrow \Omega ^{\prime }$,
if and only if $\Omega ^{\prime }\left( x^{\prime },y^{\prime }\right)
=\Omega \left( x,y\right) .~$The later conditions are expressed in terms of
the infinitesimal transformation (\ref{PT.02}) as $L_{\xi }\Omega =0$.

However, in terms of geometric objects the concept of symmetry has been
generalized, such that a geometric $\Omega $ object under the action of a
point transformation to differs by a tensor $\Psi $, that is, the
generalized conditions $\mathcal{L}_{X}\Omega =\Psi .$ While $\Omega $ can
be any geometric object with arbitrary transformation law, the Lie
derivative $\mathcal{L}_{X}\Omega $ has the same transformation law with a
tensor which means that $\Psi $ is always a tensor field. Vector fields $X$
which satisfy such relations are called geometric collineations. \ In the
case of Riemannian manifolds, the most common collineations are these of the
metric tensor and of the connection coefficients.

\subsection{Collineations of the metric tensor}

Assume a Riemannian space $V^{n}$ of dimension $n,~\dim V^{n}=0$, with
metric tensor $g_{ij}$ and connection coefficient $\Gamma _{jk}^{i}$. \ Then
a vector field $X$ is characterized as a\ Conformal Killing vector (CKV) for
the Riemannian space if
\begin{equation}
\mathcal{L}_{X}g_{ij}=2\psi g_{ij}~,~\psi =\frac{1}{n}X_{;k}^{k}.
\end{equation}

CKVs have the property to keep invariant the angle between two directions at
a point, under the action of the point transformations. The CKVs\ of a given
metric tensor $g_{ij}$ form a Lie algebra known as conformal algebra for the
space $V^{n}$. There are some important properties for the conformal
algebra. For instance, conformal related metrics, i.e. $\bar{g}_{ij},~g_{ij}$
with $\bar{g}_{ij}=Ng_{ij}$ admit the same conformal algebra, while the
maximum dimension of the admitted conformal algebra can be $\frac{1}{2}%
\left( n+1\right) \left( n+2\right) $. In that case the space with line
element $g_{ij}$ is characterized as conformally flat and there exists a
coordinate system where it can be written as $g_{ij}=N\eta _{ij}$, where $%
\eta _{ij}$ notes the metric tensor for the flat space.

In the special case where function $\psi $ is constant, that is, $\psi
_{,i}=0$, the CKV is reduced to the Homothetic Killing vector (HKV). A
Riemannian space can admits at maximum one proper HKV. Finally, when $\psi
=0 $, the vector field $X$ is called a Killing vector field or isometry.

Isometries are the most important symmetries in geometries because they keep
invariant the distances and the angles in a Riemannian space. Most important
isometries are the translations and the rotations of Euclidean space.
Because of these symmetries the geometric objects do not change when they
rotate or change location in the physical space.

The KVs form a Lie algebra known as Killing algebra of maximum dimension $%
\frac{1}{2}n\left( n+1\right) $. In the later case, the space $g_{ij}$ is
maximally symmetric and it is that of the flat space or of the $n$%
-dimensional sphere or of the $n$-dimensional hyperbolic plane. If the
maximally symmetric space admits a (proper) HKV then it is a flat space.
Maximally symmetric spaces are conformally flat which means that they admit
the maximum conformal Lie algebra of dimension$~\frac{1}{2}\left( n+1\right)
\left( n+2\right) .$

\subsection{Collineations of the connection}

Let us assume the Levi-Civita connection $\Gamma _{jk}^{i}$, then by
definition, the Lie derivative with respect to the field $X$ is written as%
\begin{equation}
\mathcal{L}_{X}\Gamma _{jk}^{i}=g^{ir}\left[ \left( \mathcal{L}%
_{X}g_{rk}\right) _{;j}+\left( \mathcal{L}_{X}g_{rj}\right) _{;k}-\left(
\mathcal{L}_{X}g_{jk}\right) _{;r}\right] .  \label{scc.01}
\end{equation}%
Hence, if $X$ is a CKV from (\ref{scc.01}) it follows
\begin{equation}
\mathcal{L}_{X}\Gamma _{jk}^{i}=2g^{ir}\left[ \psi _{;j}+\psi _{;k}-\psi
_{;r}\right] .
\end{equation}

Therefore, if $X$ is a HKV or a KV, it follows $\mathcal{L}_{X}\Gamma
_{jk}^{i}=0$, which means that the Levi-Civita connection remain invariant
under the action of Killing symmetry and of Homothetic symmetry. However,
that is not true for a proper Conformal symmetry. The collineations of the
connection can be defined independently from the collineations of the metric
tensor.

Consider now the general connection $\Gamma _{jk}^{i}$, then the vector
field $X$ is an Affine collineation (AC) if
\begin{equation}
L_{\xi }\Gamma _{jk}^{i}=0,  \label{scc.02}
\end{equation}%
This family of point transformations caries geodesic trajectories into
geodesic trajectories while preserve the affine parameter of the geodesic
equations. ACs form the so-called Affine Lie algebra and in the case of the
Levi-Civita connection the Affine Lie algebra has the subalgebras the
Homothetic algebra and the Killing algebra. The maximum dimension of the
Affine algebra is $n\left( n+1\right) $ which is that of the flat space.

Another family of collineations for the connection of special interest are
the Projective collineations (PC). Projective transformations transform the
system of geodesics (auto parallel curves) of $V^{n}$ into the same system
but they do not preserve the affine parameter. A vector field $X$ is called
a PC if there exists an one-form$~\omega _{i}$ where
\begin{equation}
L_{\xi }\Gamma _{jk}^{i}=\omega _{j}\delta _{k}^{i}+\omega _{k}\delta
_{j}^{i}.
\end{equation}

However, in the case of Riemannian manifolds the one-form~$\omega _{i}$ is
always closed, i.e. $\omega _{i}=\phi _{,i}$ which means that the later
symmetry condition is equivalent to
\begin{equation}
L_{\xi }\Gamma _{jk}^{i}=\phi _{,j}\delta _{k}^{i}+\phi _{,k}\delta _{j}^{i}.
\label{scc.03}
\end{equation}%
Function $\phi $ is called the projective function. When $\phi $ vanishes
the PC reduces to that of AC while when $\phi _{;ij}=0$ the PC is
characterized as special. The maximum dimension of the projective algebra is
$n\left( n+2\right) $, and it is that of the maximal symmetric spaces \cite%
{bar1}.

As far as the existence of a special PC is concerned, there are some
important results in the literature which are necessary for our study \cite%
{bar1}. Indeed, if the Riemannian space admits a $p\leq n$ dimensional Lie
algebra of special PCs then also admits $p$ gradient KVs and a gradient HV
and if $p=n$ the space is flat, the reverse also holds true. Furthermore, a
maximally symmetric space which admits a proper AC or a special PC is a flat
space.

There is a zoology of definitions for collineations in\ Riemannian spaces,
we refer the reader in the interesting diagram in \cite{bar3}.

At this point it is important to mention that with the term gradient
collineation we refer to a collineation vector field $X$, which can be
written as a closed one-form. Moreover, with the term proper collineation we
refer to a collineation with a specific property. For instance as a proper
AC we refer to an AC which is not a KV or a HKV, while with the term proper
PC we refer to a collineation which is a PC and not a AC or HKV or KV.

\section{Symmetries of ordinary differential equations}

\label{sec4}

Below the application of point symmetries on differential equations is
presented. We give the definition of a Lie point symmetry for differential
equations and we show how the Lie invariants are applied to simplify a given
differential equation.

A differential equation $H=0$ is a function defined in the jet space $%
B_{M}=\left\{ x,y,\dot{y},\ddot{y}...,y^{\left( n\right) }\right\} $, where
we have assumed $x$ to be the independent variable and $y=y\left( x\right) $
to be the dependent variable and $\dot{y}=\frac{dy}{dx}~,~\ddot{y}=\frac{%
d^{2}y}{dx^{2}}~,~...~$,~$y^{\left( n\right) }=\frac{d^{\left( n\right) }y}{%
dx^{n}}$. Therefore function $H$ is defined as $H=H\left( x,y,\dot{y},\ddot{y%
}...,y^{\left( n\right) }\right) \,$.

Consider now the infinitesimal transformation in the basic manifold $\left\{
\bar{x},\bar{y}\right\} \rightarrow \left\{ x+\varepsilon \xi \left(
x,y\right) ,y+\varepsilon \eta \left( x,y\right) \right\} $, where in the
jet space $B_{M}$ the point transformation is prolonged as
\begin{eqnarray*}
\bar{x} &=&x+\varepsilon \xi \left( x,y\right) \\
\bar{y} &=&x+\varepsilon \eta \left( x,y\right) \\
\bar{y}^{\left( 1\right) } &=&y^{\left( 1\right) }+\varepsilon \eta ^{\left[
1\right] } \\
&&... \\
\bar{y}^{\left( n\right) } &=&y^{\left( n\right) }+\varepsilon \eta ^{\left[
n\right] }
\end{eqnarray*}%
with generator vector%
\begin{equation}
X^{\left[ n\right] }=\xi \partial _{x}+\eta \partial _{\eta }+\eta ^{\left[ 1%
\right] }\partial _{\dot{y}}+...+\eta ^{\left[ n\right] }\partial
_{y^{\left( n\right) }}\text{.}  \label{scc.04}
\end{equation}

Functions $\eta ^{\left[ n\right] }$ are known as the
prolongation/extensions functions and they are defined as%
\begin{eqnarray}
\eta ^{\left[ 1\right] } &=&\dot{\eta}-y^{\left( 1\right) }\dot{\xi}, \\
\eta ^{\left[ 2\right] } &=&\dot{\eta}^{\left[ 1\right] }-\ddot{y}\dot{\xi},
\\
&&...  \notag \\
\eta ^{\left[ n\right] } &=&\frac{d\eta ^{n-1}}{dx}-y^{\left( n\right) }%
\frac{d\xi }{dx}=\frac{d^{n}}{dx^{n}}\left( \eta -y^{\left( 1\right) }\xi
\right) +y^{\left( n+1\right) }\xi .
\end{eqnarray}

Therefore, the differential equation $H\left( x,y,\dot{y},\ddot{y}%
...,y^{\left( n\right) }\right) =0$ is invariant under the action of an
parameter point transformation if the following condition is true%
\begin{equation}
X^{\left[ n\right] }\left( H\left( x,y,\dot{y},\ddot{y}...,y^{\left(
n\right) }\right) \right) =0,  \label{scc.05}
\end{equation}%
or equivalently
\begin{equation}
X^{\left[ n\right] }\left( H\left( x,y,\dot{y},\ddot{y}...,y^{\left(
n\right) }\right) \right) =\lambda \left( H\left( x,y,\dot{y},\ddot{y}%
...,y^{\left( n\right) }\right) \right) ~,~\text{with }H=0\text{.}
\label{scc.06}
\end{equation}

Hence, when the symmetry condition (\ref{scc.05}) is true, the infinitesimal
generator $X$ will be called Lie point symmetry for the differential
equation. The basic application of the Lie point symmetries are summarized
in the application of the similarity transformations which are used to
reduce the order on differential equations or the number of dependent
variables for partial differential equations.

There are different ways to apply the Lie point symmetries for the reduction
of a given differential equation.\ However, the different approaches are
equivalent. The most common methods are the derivation of canonical
transformations, and the derivation of the Lie invariants. \

Let $X=\xi \left( x,y\right) \partial _{x}+\eta \left( x,y\right) \partial
_{y}$ $\ $be a Lie point symmetry for the differential equation $H\left( x,y,%
\dot{y},\ddot{y}...,y^{\left( n\right) }\right) \equiv y^{\left( n\right)
}-\omega \left( x,y,\dot{y},...,y^{\left( n-1\right) }\right) =0$. Then
under the change of variables $\left\{ x,y\right\} \rightarrow \left\{
r,s\right\} $ such that
\begin{equation}
Xr=0~~,~~Xs=1  \label{cc.00}
\end{equation}%
the symmetry vector reads $X=\partial _{s}$, while the differential equation
as%
\begin{equation}
\frac{d^{n}s}{dr^{2}}=\bar{\omega}\left( r,s,\frac{ds}{dr},...,\frac{d^{n-1}s%
}{dr^{n-1}}\right) .  \label{cc.00a}
\end{equation}

By definition it follows $\frac{\partial }{\partial s}\bar{\omega}\left( r,s,%
\frac{ds}{dr},...,\frac{d^{n-1}s}{dr^{n-1}}\right) =0,$ therefore by define
the new variable $S=\frac{ds}{dr}$, we are able to reduce the order of the
differential equation\ and rewrite it as%
\begin{equation}
\frac{d^{n-1}S}{dr^{n-1}}=\bar{\omega}\left( r,S,\frac{dS}{dr},...,\frac{%
d^{n-2}S}{dr}\right) .  \label{cc.00c}
\end{equation}

The second approach that we discuss for the application of the Lie point
symmetries for the reduction of a differential equation is based on the
derivation of the differential invariant functions which follows from the
Lagrange system%
\begin{equation}
\frac{dx}{\xi }=\frac{dy}{\eta }=\frac{d\dot{y}}{\eta _{\left[ 1\right] }}%
=...=\frac{dy^{\left( n\right) }}{\eta _{\left[ n\right] }}.  \label{Ip.02}
\end{equation}

The system (\ref{Ip.02}) provides us with characteristic functions
\begin{equation*}
W^{\left[ 0\right] }\left( x,y\right) ,~W^{\left[ 1\right] }\left( x,y,\dot{y%
}\right) ,~W^{\left[ n\right] }\left( x,y,\dot{y},\ddot{y},...,y^{\left(
n\right) }\right)
\end{equation*}%
where $W^{\left[ n\right] }$ is the nth order invariant of the Lie symmetry
vector. By considering \ as $u=W^{\left[ 0\right] }$ to be the new
independent variable and $v=W^{\left[ 1\right] }$ the new dependent variable
then the differential equation $H\left( x,y,\dot{y},\ddot{y}...,y^{\left(
n\right) }\right) \equiv y^{\left( n\right) }-\omega \left( x,y,\dot{y}%
,...,y^{\left( n-1\right) }\right) =0$ can be written in the equivalent form

Let $u=W^{\left[ 0\right] }~,~v=W^{\left[ 1\right] },$ where $W^{\left[ 0%
\right] },~W^{\left[ 1\right] }$ are the zero and the first order invariants
of a Lie symmetry repetitively. From zero-order and first-order invariants $%
u,~v,~$the higher-order differential invariants
\begin{equation}
\frac{dv}{du}~,...,\frac{d^{n-1}v}{du^{n-1}},  \label{Ip.04}
\end{equation}%
can be defined.

The differential invariants are functions of the derivatives $y^{\left(
n\right) }$. Hence, the given differential given $H\left( x,y,\dot{y},\ddot{y%
}...,y^{\left( n\right) }\right) =0,$ may be written in terms of the
differential invariants (\ref{Ip.04}), i.e.%
\begin{equation}
\frac{d^{n-1}v}{du^{n-1}}=\Omega \left( u,v,\frac{dv}{du}~,...\frac{d^{n-1}v%
}{du^{n-1}}\right) ,
\end{equation}%
where $\frac{dv}{du}=\frac{\frac{\partial v}{\partial x}+\frac{\partial v}{%
\partial y}\dot{y}+...+\frac{\partial v}{\partial \dot{y}}\ddot{y}}{\frac{%
\partial v}{\partial x}+\frac{\partial v}{\partial y}\dot{y}}.~$

There are alternative methods and approaches to apply Lie symmetries. A
quite intriguing application of Lie symmetries is to produce integrals or
Lagrangian functions for a system of ODEs by the \ method of Jacobi's last
multiplier,see for instance \cite{nuc1}, while there are very interesting
approaches for the construction of conservation laws by using the Lie point
symmetries \cite{nuc2,nuc3}.

The most famous approach for the derivation of conservation laws from the
point symmetries for systems with an Action principle is described by
Noether's theorems. Noether's theorem for second-order differential
equations $H\left( x,y,\dot{y},\ddot{y}\right) =0~$is presented.

Let function~$L=L\left( x,y,\dot{y}\right) $ be the Lagrangian for the
differential equation $H\left( x,y,\dot{y},\ddot{y}\right) =0$, that is, $%
E_{L}\left( L\right) =H\left( x,y,\dot{y},\ddot{y}\right) $, where $E_{L}$
is the Euler-Lagrange vector. Therefore, if the following condition is true%
\begin{equation}
X^{\left[ 1\right] }L+L\frac{d\xi }{dx}=\frac{df}{dx}  \label{Ns.03}
\end{equation}%
where $f$ is a function, then the variation of the Action integral is
invariant. The later it means that $X$ is a Lie point symmetry for the
differential equation $H\left( x,y,\dot{y},\ddot{y}\right) =0$, while$~X$ is
called a Noether symmetry for the Lagrangian function $L\left( x,y,\dot{y}%
\right) $.

Moreover, from Noether's second theorem it follows that for every Noether
symmetries $X$ the following function is a conservation laws for the
differential equation
\begin{equation}
\Phi \left( x,y,\dot{y}\right) =\xi \left( x,y\right) \left( \dot{y}\frac{%
\partial L}{\partial \dot{y}}-L\right) -\eta \left( x,y\right) \frac{%
\partial L}{\partial y}+f,  \label{Ns.04a}
\end{equation}%
that is, $\frac{d\Phi }{dx}=0$.

Noether symmetries for a given dynamical system form a Lie algebra known as
Noether algebra. The Noether algebra is a subalgebra of the admitted Lie
algebra for the dynamical system.

\section{Symmetries of Geodesic equations}

\label{sec5}

We continue our analysis by studying the Lie point symmetries for the
geodesic equations. In particular, we review previous results which connect
the Lie point symmetries of geodesic trajectories with the collineations of
the background space. Because the following applications are of special
interests in physics $t~$~is assumed to be the independent variable and $%
x^{i}=x^{i}\left( t\right) $ to be the dependent variables.

Let as assume the $C^{p}$ manifold of dimension $n$ with the connection $%
\Gamma _{jk}^{i}.~$Thus, in a local coordinate system the autoparallels are
defined as
\begin{equation}
\ddot{x}^{i}+\Gamma _{jk}^{i}\dot{x}^{j}\dot{x}^{k}+\phi \left( t\right)
\dot{x}^{i}=0  \label{PCA.1}
\end{equation}%
where $t$ is a parameter along the paths, and a dot means derivative with
respect to s. When $\phi \left( t\right) $ vanishes, the autoparallels are
affine parametrized and $t$ is the affine parameter.

Without loss of generality a new variable $dt=f\left( t\right) dt$ can
always be defined such that equations (\ref{PCA.1}) to be written as%
\begin{equation}
\frac{d^{2}x^{i}}{dS^{2}}+\Gamma _{jk}^{i}\frac{dx^{j}}{dS}\frac{dx^{k}}{dS}%
+\left( \frac{dS}{dt}\right) ^{-2}\left( \left( \frac{d^{2}S}{dt^{2}}\right)
+\phi \left( S\right) \left( \frac{dS}{ds}\right) \right) \frac{dx^{i}}{dS}=0
\end{equation}%
or
\begin{equation}
\frac{d^{2}x^{i}}{dS^{2}}+\Gamma _{jk}^{i}\frac{dx^{j}}{dS}\frac{dx^{k}}{dS}%
=0
\end{equation}%
where we have set $\frac{df\left( t\right) }{dt}+\phi \left( t\right)
f\left( t\right) =0$. Therefore in the following it is assumed $\phi \left(
t\right) =0$.

Consider the infinitesimal transformation

\begin{equation}
\bar{t}=s+\varepsilon \xi \left( s,x^{k}\right) ~,~~\bar{x}%
^{i}=x^{i}+\varepsilon \eta ^{i}\left( t,x^{k}\right)  \label{PCA.1a}
\end{equation}%
with generator $X=\xi \left( t,x^{k}\right) \partial _{t}+\eta ^{i}\left(
t,x^{k}\right) \partial _{i}.$\

The autoparallels (\ref{PCA.1}) are invariant under the transformation (\ref%
{PCA.1a}) iff the following conditions holds
\begin{equation}
X^{\left[ 2\right] }\left( \ddot{x}^{i}+\Gamma _{jk}^{i}\dot{x}^{j}\dot{x}%
^{k}\right) =0\;  \label{PCA.1C}
\end{equation}%
by replacing $X^{\left[ 2\right] }$ from its definition and consider the
coefficients of polynomial of the derivatives of $x^{i}$ to be equal with
zero follows the following system
\begin{equation}
\eta _{,tt}^{i}=0  \label{PCA.31}
\end{equation}%
\begin{equation}
\xi _{,tt}\delta _{j}^{i}-2\left[ \eta _{,tj}^{i}+\eta _{,t}^{k}\Gamma
_{(kj)}^{i}\right] =0  \label{PCA.32}
\end{equation}

\begin{equation}
\mathcal{L}_{\mathbf{\eta }}\Gamma _{(jk)}^{i}=\xi \Gamma _{(kj),t}^{i}+2\xi
_{,t(j}\delta _{k)}^{i}  \label{PCA.34}
\end{equation}%
\begin{equation}
\xi _{(,j|k}\delta _{d)}^{i}=0\;.  \label{PCA.35}
\end{equation}

We proceed with the solution of the symmetry conditions (\ref{PCA.31})-(\ref%
{PCA.35}). In addition we consider that the connection coefficients they do
not depend on the indepedent parameter $t$.

From equations (\ref{PCA.31}) and (\ref{PCA.35}) we derive $\eta
^{i}(t,x^{k})=A^{i}(x^{k})t+B^{i}(x^{k})$ and $\xi
(t,x)=C_{J}(t)S^{J}(x^{k})+D(t)$. Functions $%
A^{i}(x^{k})~,~B^{i}(x^{k})~,~C_{J}(t)~,~D(t)$ are arbitrary functions which
will be constraint by the symmetry conditions (\ref{PCA.32}), (\ref{PCA.34})
while $S^{J}(x^{k})$ is the generating function for a gradient KV, i.e. $%
S^{J}(x)_{|(i,j)}=0$. Index $J$ run on the number of independent gradient
KVs.

Furthermore, by replacing in (\ref{PCA.32}) it follows%
\begin{equation}
2A(x^{k})_{;j}^{i}=\left[ C_{J}(t),_{tt}S^{J}(x^{k})+D(t),_{tt}\right]
\delta _{j}^{i},  \label{sc.01}
\end{equation}%
that is;
\begin{eqnarray}
D(t) &=&\frac{1}{2}Mt^{2}+Kt+L\,, \\
C_{J}(t) &=&\frac{1}{2}G_{J}t^{2}+E_{J}t+F_{J},  \notag
\end{eqnarray}%
in which~$M~,$ $K~,~L~$,~ $G_{J}~,$ $E_{J}~,$ $F_{J}$ are constants.

In addition, for the function~$A^{i}(x^{k})$, from (\ref{sc.01}) we derive $%
A(x^{k})_{i;j}=\frac{1}{2}\left( G_{J}S^{J}(x^{k})+M\right) g_{ij}$, where $%
g_{ij}$ is the background metric tensor. Consequently, $A^{i}\left(
x^{k}\right) $ is an element of the conformal algebra for the metric tensor $%
g_{ij}$. However, \ antisymmetric part vanishes, i.e. $A\left( x^{k}\right)
_{\left[ i;j\right] }=0$, which means that~\thinspace $A^{i}\left(
x^{k}\right) $ is a gradient CKV with conformal factor~$\psi =\frac{1}{2}%
(G_{J}S^{J}(x)+M)$.

Finally, from the symmetry condition (\ref{PCA.34}) we end with the
following system
\begin{eqnarray}
L_{A}\Gamma _{jk}^{i} &=&2G_{J}S^{J}(x)_{,(j}\delta _{k)}^{i}, \\
L_{B}\Gamma _{jk}^{i} &=&2E_{J}S^{J}(x)_{,(j}\delta _{k)}^{i},
\end{eqnarray}%
that is, the vector fields $A^{i}(x^{k}),B^{i}(x^{k})$ are special PCs with
projective functions $G_{J}S^{J}(x)$ and $E_{J}S^{J}(x)$ respectively.
However, because $A^{i}\left( x^{k}\right) $ is also an element of the
conformal group for the underlying metric, it follows that $A^{i}\left(
x^{k}\right) $ is a gradient KV or a gradient HV.

We conclude that the Lie point symmetries for the geodesic equations, or for
the autoparallels, are constructed by the elements of the special projective
algebra of the background space as described by the following theorem

\begin{theorem}
\label{TheoremGEs} The generic Lie point symmetry vector $X=\xi
(t,x^{k})\partial _{t}+\eta ^{i}(t,x^{k})\partial _{i}~$for the geodesic
equations
\begin{equation}
\ddot{x}^{i}+\Gamma _{jk}^{i}\left( x^{k}\right) \dot{x}^{j}\dot{x}^{k}=0
\end{equation}%
in a Riemannian background space with metric tensor $g_{ij}\left(
x^{k}\right) $, where $\Gamma _{jk}^{i}\left( x^{k}\right) $ is the
Levi-Civita connection, is generated by the elements of the special
projective algebra for the Riemannian manifold.

When the background space admits gradient KVs $S^{J}\left( x^{k}\right) $,
functions $\xi (t,x^{k})$,~$\eta ^{i}(t,x^{k})$ are given by the formula%
\begin{equation*}
\xi (t,x^{k})=\frac{1}{2}\left( G_{J}S^{J}\left( x^{k}\right) +M\right)
t^{2}+\left[ E_{J}S^{J}\left( x^{k}\right) +K\right] t+F_{J}S^{J}\left(
x^{k}\right) +L,
\end{equation*}%
\begin{equation}
\eta ^{i}(t,x^{k})=A^{i}(x^{k})t+B^{i}(x^{k})+D^{i}(x^{k})
\end{equation}%
where $G_{J},M,b,K,F_{J}$ and $L$ \ are constants and the index $J\ $runs
along the number of gradient KVs, $A^{i}(x)$ is a gradient HV with conformal
factor $\psi =\frac{1}{2}\left( G_{J}S^{J}\left( x^{k}\right) +M\right) $, $%
D^{i}(x)$ is a non-gradient KV of the metric and $B^{i}(x)$ is either a
special projective collineation with projection function $E_{J}S^{J}(x)$ or
an AC and $E_{J}=0$.

When the background space does not admit gradient KVs, functions $\xi
(t,x^{k})$,~$\eta ^{i}(t,x^{k})$ are given by the formula%
\begin{equation}
\xi (t,x^{k})=\frac{1}{2}Mt^{2}+Kt+L
\end{equation}%
\begin{equation}
\eta ^{i}(t,x^{k})=A^{i}(x^{k})t+B^{i}(x^{k})+D^{i}(x^{k}),
\end{equation}%
where $A^{i}(x)$ is a gradient HV with conformal factor $\psi =\frac{1}{2}M,$
$D^{i}(x)$ is a non-gradient KV\ of the metric and $B^{i}(x)$ is an AC.

Furthermore, when the metric tensor does not admit gradient KV and gradient
HKV, functions $\xi (t,x^{k})$,~$\eta ^{i}(t,x^{k})$ are given by the
formula
\begin{align}
\xi (t)& =Kt+L \\
\eta ^{i}(x^{k})& =B^{i}(x^{k})+D^{i}(x^{k}).
\end{align}%
Finally, if the background vector field admits a zero-dimensional special
projective algebra the generic Lie point symmetry is $X=\left( Kt+L\right)
\partial _{t}$.
\end{theorem}

We proceed our discussion with the investigation of the Noether symmetries
for the geodesic equations. Because Noether symmetries form a subalgebra of
the Lie symmetries, Noether vector fields are constructed by the elements of
the special projective algebra for the background space.

Consider the geodesic Lagrangian function
\begin{equation}
L\left( x^{k},\dot{x}^{k}\right) =\frac{1}{2}g_{ij}\left( x^{k}\right) \dot{x%
}^{i}\dot{x}^{j}~.  \label{NS.6}
\end{equation}%
Then, from the Noether symmetry condition (\ref{Ns.03}) for the
infinitesimal generator $X=\xi \left( t,x^{k}\right) \partial _{t}+\eta
^{i}\left( t,x^{k}\right) \partial _{i}~$we end with the following system of
partial differential equations
\begin{align}
\xi _{,k}& =0  \label{NS.8} \\
\mathcal{L}_{\eta }g_{ij}& =2\left( \frac{1}{2}\xi _{,t}\right) g_{ij}
\label{NS.9} \\
\eta _{,t}^{,i}g_{ij}& =f_{,i}  \label{NS.10} \\
f_{,t}& =0  \label{NS.11}
\end{align}

From (\ref{NS.11})\thinspace\ it follows $\xi \left( t,x^{k}\right) =\xi
\left( t\right) $, while condition (\ref{NS.10}) gives $\eta _{i}\left(
t,x\right) =f_{,i}\left( t,x^{k}\right) t+K_{i}(x^{j}).$ Additionally,
equation (\ref{NS.9}) provides that $\xi _{,t}=2\psi $, where $\psi =const.$
while $\eta ^{i}\left( t,x^{k}\right) $ is HKV for the background space. By
following the same steps with as in the case of Lie point symmetries, for
the Noether point symmetries for the geodesic Lagrangian our results are
summarized in the following theorem

\begin{theorem}
\label{TheoremGE2}The Noether Symmetries of the geodesic Lagrangian (\ref%
{NS.6}) are generated by the KVs and the HKV of the metric $g_{ij}$ as
follows:%
\begin{eqnarray}
X &=&\left( C_{3}\psi t^{2}+2C_{2}\psi t+C_{1}\right) \partial _{t}+  \notag
\\
&&+\left[ C_{J}S^{J,i}+C_{I}KV^{Ii}+C_{IJ}tS^{J,i}+C_{2}H^{i}+C_{3}t(GHV)^{i}%
\right] \partial _{i}  \label{NS.14a}
\end{eqnarray}%
with corresponding gauge function%
\begin{equation}
f(x^{k})=C_{1}+C_{2}+C_{I}+C_{J}+\left[ C_{IJ}S^{J}\right] +C_{3}\left[ GHV%
\right] ,  \label{NS.15}
\end{equation}%
where $S^{J,i}$ are the $C_{J}$ gradient KVs, $KV^{Ii}$ are the $C_{I}$
non-gradient KVs, $H^{i}$ is a HKV\ not necessarily gradient~$\left[ GHV%
\right] $ and is the gradient HKV (if it exists) of the metric $g_{ij}$.
\end{theorem}

Finally, from Noether's second theorem and Theorem \ref{TheoremGE2} for the
conservation laws of the geodesic equations it follows the theorem.

\begin{theorem}
The generic form of the Noetherian conservation law for the geodesic
Lagrangian (\ref{NS.6}) is
\begin{align}
\phi & =\frac{1}{2}\left[ C_{3}\psi t^{2}+2C_{2}\psi t+C_{1}\right] g_{ij}%
\dot{x}^{i}\dot{x}^{j}  \notag \\
& +\left[
C_{J}S^{J,i}+C_{I}KV^{Ii}+C_{IJ}tS^{J,i}+C_{2}H^{i}(x^{r})+C_{3}t(GHV)^{,i}%
\right] g_{ij}\dot{x}^{j}  \notag \\
& +C_{1}+C_{2}+C_{I}+C_{J}+\left[ C_{IJ}S^{J}\right] +C_{3}\left[ GHV\right]
.
\end{align}%
The individual conservation laws linear in the momentum are%
\begin{eqnarray}
C_{I} &\neq &0:\phi _{C_{I}}=KV_{i}^{I}\dot{x}^{i}-1~, \\
C_{J} &\neq &0:g_{ij}S^{J,i}\dot{x}^{j}-1~, \\
C_{IJ} &\neq &0:tg_{ij}S^{J,i}\dot{x}^{j}-S^{J}~,
\end{eqnarray}%
while the quadratic in the momentum conservation laws are%
\begin{eqnarray}
C_{1} &\neq &0:\phi _{C_{1}}=\frac{1}{2}g_{ij}\dot{x}^{i}\dot{x}^{j}~+1, \\
C_{2} &\neq &0:\phi _{C_{2}}=t\psi g_{ij}\dot{x}^{i}\dot{x}^{j}-g_{ij}H^{i}%
\dot{x}^{j}+1~, \\
C_{3} &\neq &0:\phi _{C_{3}}=\frac{1}{2}t^{2}\psi g_{ij}\dot{x}^{i}\dot{x}%
^{j}-t(GHV)_{,i}\dot{x}^{i}+\left[ GHV\right] .
\end{eqnarray}
\end{theorem}

From Theorems \ref{TheoremGEs} and \ref{TheoremGE2}\ it is clear that one is
able to compute the Lie symmetries and the Noether symmetries of the
geodesic equations in Riemannian manifolds by derive the collineation
vectors and avoid the cumbersome formulation of the Lie symmetry method.
However, the inverse procedure is true. In the following, we focus on the
inverse approach in order to construct projective algebra of decomposable
spacetimes by the Lie point symmetries of the nondecomposable part of the
space.

\section{Projective collineations of decomposable spacetimes}

\label{sec6}

The symmetry condition (\ref{PCA.34}) can be written in the equivalent form%
\begin{equation}
\mathcal{L}_{\eta }\Gamma _{(jk)}^{i}+\mathcal{L}_{\xi }\Gamma
_{(jk)}^{i}=2\Phi _{(,j}\delta _{k)}^{i}
\end{equation}%
in which $\Phi =\xi _{,t}.\,$Furthermore, if we sum the symmetry conditions (%
\ref{PCA.31})-(\ref{PCA.35}) we end with the equation%
\begin{equation}
\mathcal{L}_{X}\Gamma _{BC}^{A}=2\Phi _{(,A}\delta _{C)}^{B}.  \label{PCA.37}
\end{equation}%
where $\Gamma _{BC}^{A}$ is the Levi-Civita connection for the decomposable
Riemannian manifold%
\begin{equation}
ds^{2}=\varepsilon dt^{2}+g_{ij}\left( x^{k}\right) dx^{i}dx^{j},
\label{dc.01}
\end{equation}%
with nonzero components $\Gamma _{BC}^{A}=\Gamma _{jk}^{i}~$in which $A,B$
are the indices in the $1+n$ spaces and $i,j$ the indices in the $n$-space.
\ Thus, the following theorem for the collineations of decomposable spaces
follows.

\begin{theorem}
\label{t3} The Lie point symmetries for the geodesic equations of the $n$%
-dimensional Riemannian space $g_{ij}\left( x^{k}\right) $ form the
projective algebra for the $\left( n+1\right) $-decomposable Riemannian
space (\ref{dc.01}) and vice versa.
\end{theorem}

From Theorem \ref{t3} and the function forms of the Lie point symmetries as
they are given in Theorem \ref{TheoremGEs}\ we end with the following
corollary{theorem}.

\begin{theorem}
For the elements of the special projective algebra of the decomposable space
(\ref{dc.01}) we have the following observations:

A) The $\left( n+1\right) $-dimensional decomposable space admits a proper
PC, then the field is a special projective collineation.

B) The $\left( n+1\right) $-dimensional decomposable space admits a HKV if
and only if the $n$-space admits a HKV.

C) The $\left( n+1\right) $-dimensional decomposable space admits a proper
PC if and only if the $n$-space admits a gradient HKV.

D) There exists a rotation among the axes $t$ and $x^{i}$ if and only if
there exists a gradient KV on the direction of $x^{i}$.

E) The $\left( n+1\right) $-dimensional decomposable space admits a
projective algebra $G^{n+1}$ of minimum dimension $2$. When $\dim G^{n+1}=2$%
, then $n$-dimensional space does not admit any special PC.
\end{theorem}

Some of the above results have been proved before in the literature by using
tools of differential geometry. However in our approach we applied tools
from the theory of symmetries of differential equations. We demonstrate the
application of Theorem \ref{t3} with some examples.

Consider the space of constant curvature
\begin{equation}
d\tau ^{2}=\frac{1}{\left( 1+\frac{1}{4}Kx^{i}x_{i}\right) ^{2}}\left(
dx^{2}+dy^{2}+dz^{2}\right)  \label{dz1}
\end{equation}%
with $K\neq 0$. The space admits a six elements on the special projective
algebra consisted by the (nongradient) Killing symmetries$~\mathbf{r}_{\mu
\nu },\mathbf{I}_{\mu }~$\cite{ts1}.\ The Lie point symmetries for the
geodesic equations of the space are the vector fields
\begin{equation*}
\mathbf{r}_{\mu \nu },\mathbf{I}_{\mu }~,\partial _{\tau }~,~\tau \partial
_{\tau }
\end{equation*}%
Consequently, for the $\left( 3+1\right) $-dimensional decomposable space%
\begin{equation}
ds^{2}=-\varepsilon d\tau ^{2}+\frac{1}{\left( 1+\frac{1}{4}%
Kx^{i}x_{i}\right) ^{2}}\left( dx^{2}+dy^{2}+dz^{2}\right)
\end{equation}%
it follows that the vector fields $\mathbf{r}_{\mu \nu },\mathbf{I}_{\mu }~$%
\ are KVs, $\partial _{\tau }$ is a gradient KV, while $\tau \partial _{\tau
}$ is an AC.

For a second application, let us assume the four-dimensional G\"{o}del
spacetime in Cartesian coordinates%
\begin{equation}
d\tau ^{2}=-dt^{2}-2e^{ax}dtdy+dx^{2}-\frac{1}{2}e^{2ax}dy^{2}+dz^{2}.
\end{equation}

The geodesic equations are%
\begin{align}
t^{\prime \prime }+2at^{\prime }x^{\prime }+ae^{ax}x^{\prime }y^{\prime }&
=0, \\
x^{\prime \prime }+ae^{ax}t^{\prime }y^{\prime }+\frac{1}{2}%
ae^{2ax}y^{\prime 2}& =0, \\
y^{\prime \prime }-2ae^{-ax}t^{\prime }x^{\prime }& =0, \\
z^{\prime \prime }& =0.
\end{align}%
The G\"{o}del metric admits the following elements for the special
projective algebra%
\begin{equation}
Y^{1}=\partial _{z}~,~Y^{3}=\partial _{x}-ay\partial _{y}~,~Y^{4}=\partial
_{t}~,~Y^{5}=\partial _{y}~,~Y^{6}=z\partial _{z}
\end{equation}%
\begin{equation}
Y^{2}=\left( -\frac{2}{a}e^{-ax}\right) \partial _{t}+y\partial _{x}+\left(
\frac{2e^{-2ar}-a^{2}y^{2}}{2a}\right) \partial _{y}
\end{equation}%
where $Y^{1}$ is a gradient KV $\left( S_{1}=z\right) $, $Y^{2-5}$ are non
gradient KVs and $Y^{6}$ is a proper AC. Consequently, the Lie point
symmetries are derived to be

\begin{equation}
X_{1}=\partial _{\tau }~~,~~X_{2}=\tau \partial _{\tau }~~,~~X_{3}=z\partial
_{\tau }~~,~~X_{4}=Y^{4}
\end{equation}%
\begin{equation}
~X_{5}=Y^{2}~~,~~X_{6}=Y^{3}~~,~~X_{7}=Y^{5}~
\end{equation}%
\begin{equation}
X_{8}=Y^{1}~,~X_{9}=\tau Y^{1}~~,~~X_{10}=Y^{6}
\end{equation}%
Thus, for the five-dimensional line element%
\begin{equation}
d\tau ^{2}=\varepsilon d\tau ^{2}-dt^{2}-2e^{ax}dtdy+dx^{2}-\frac{1}{2}%
e^{2ax}dy^{2}+dz^{2}
\end{equation}%
vector field$~X_{1}$ is a gradient KV, $X_{2},~X_{3}\,\ \,\,$and $~X_{9}$
are proper ACs, $X_{4-10}$ have the same properties as that for the
four-dimensional manifold. We observe that the field $X_{3}-\varepsilon
X_{9} $ is a rotation on the directions $\tau -z$, which is a Killing
symmetry for the five-dimensional space.

As a final example consider the three-dimensional line element
\begin{equation}
dx_{\left( 3\right) }^{2}=-dt^{2}+\left( at+b\right) ^{2}\left(
dy^{2}+dz^{2}\right) .  \label{sn1.}
\end{equation}%
which admits the three nongradient KVs
\begin{equation}
K^{1}=\partial _{y}~,~K^{2}=\partial _{z}~,~K^{3}=z\partial _{y}-y\partial
_{z}.
\end{equation}%
and the gradient HKV
\begin{equation}
H^{i}=\left( t+\frac{b}{a}\right) \partial _{t}~,~\psi =1.
\end{equation}

Thus, the Lie point symmetries for the geodesic equations for the space are
the vector fields%
\begin{equation}
\partial _{x},~x\partial _{x}~,~K^{1},~K^{2},~K^{3}~,~H^{i}~,~~x^{2}\partial
_{x}+xH^{i}.
\end{equation}%
Hence, for the four-dimensional space%
\begin{equation}
ds^{2}=-dt^{2}+dx^{2}+\left( at+b\right) ^{2}\left( dy^{2}+dz^{2}\right)
\end{equation}%
it follows that $K^{1},~K^{2},~K^{3}$are nongradient KVs,~$\partial _{x}$~is
a gradient KV, $x\partial _{x}+H^{,i}~$is the gradient HKV; $x\partial
_{x}~,~H^{i}$ are ACs. Finally, $x^{2}\partial _{x}+xH^{i}~$is a proper
special\ PC for the decomposable spacetime.

\section{Conclusions}

\label{sec7}

The theory of symmetries for differential equations is important for the
study of the integrability properties and the determination of conservation
laws. The later can be used to understand the trajectories of geometrical
objects such are the orbits of astrophysical objects \cite{ar1,ar3}.
Furthermore, the procedure for the derivation of the Lie symmetries it is
straightforward, but usually it is a high dimension system and symbolic
computation software is usually applied \cite{ar4}. In this study we
reviewed previous published results for the geometric description and
construction of the Lie point symmetries. This approach connects the two
different studies between the geometric properties of Riemannian manifolds
and of the symmetries for differential equations. We show that important
results of differential geometry can be derived easily by using the Lie
symmetry analysis.

In this spirit, we focused on the geometric interpretation of the Lie point
symmetries for the geodesic equations. We proved that the Lie point
symmetries are the elements of the projective algebra for an extended
decomposable space, while the projective algebra is directed related with
the special projective algebra of the nondecomposable space. The existence
criteria for the nature of specific geometric collineations, such are the
rotations, the proper ACs, or the proper special PC discussed in details.

This study contributes in the subject for the geometrization of the
symmetries of differential equations. Such geometric approach has been
applied also in the case of partial differential equations where different
relations between the Lie point symmetries and the collineations of the
background space were found \cite{ar5,ar6,ar7}. This geometric approach can
be seen as a connection bridge between the differential geometry and the
theories of differential equations, and applied mathematics. In this study
we show an alternative way for the study of existence theorems, in the
future we plan to investigate further applications for this geometric point
of view.

\end{document}